\documentclass[twocolumn]{svjour3}          

\usepackage{graphicx}                       
\usepackage{amsmath,amsfonts,amssymb}       
\usepackage{hyperref}                       
\usepackage{float}                          
\usepackage{lipsum}                         
\usepackage{newtxtext}                      
\usepackage{newtxmath}                      
\usepackage{multirow}                       
\usepackage[title]{appendix}                
\usepackage{xcolor}                         
\usepackage{textcomp}                       
\usepackage{booktabs}                       
\usepackage{algorithm}                      
\usepackage{algpseudocode}                  
\usepackage{listings}                       

\usepackage{geometry}
\usepackage{booktabs}    
\usepackage{tabularx}    
\usepackage{array}       
\usepackage{ragged2e}    
\usepackage{seqsplit}    
\usepackage{makecell}    
\usepackage[T1]{fontenc} 
\usepackage{xcolor}      

\usepackage{titlesec}

\titlespacing*{\section}
  {0pt}{2.8ex plus 1.4ex minus .7ex}{2.0ex plus .8ex}

\titlespacing*{\subsection}
  {0pt}{2.5ex plus 1.2ex minus .6ex}{1.8ex plus .6ex}

\titlespacing*{\subsubsection}
  {0pt}{2.2ex plus 1.0ex minus .5ex}{1.5ex plus .5ex}

\renewcommand{\arraystretch}{1.1}  
\newcolumntype{Y}{>{\RaggedRight\arraybackslash}X}



\journalname{Springer Journal}              

\title{AI-Powered Multi-Stakeholder Ecosystems for Global Development: A Design Research Study on the GSI D-Hub Proof-of-Concept Platform}

\author{Muzakkiruddin Ahmed Mohammed \and Adeeba Tarannum \and Eileen Devereux Dailey \and Marla Johnson \and Mert Can Cakmak \and John Talburt }
\institute{
\email{mmohammed6@ualr.edu, atarannum@ualr.edu, eddailey@ualr.edu, mkjohnson@ualr.edu, mccakmak@ualr.edu, jrtalburt@ualr.edu}
Center for Entity Resolution and Information Quality (ERIQ) - University of Arkansas - Little Rock, Arkansas, Little Rock, USA
*Corresponding author. E-mail: \href{mailto:iauthor@gmail.com}{mccakmak@ualr.edu}. \\
}

\date{}
\begin{document}

\maketitle

\begin{abstract}

Digital platforms increasingly support collaboration across organizations, yet many remain constrained by fragmented data and limited transparency. This paper presents the Global Solutions Initiative (GSI) D-Hub, a data-driven coordination platform that applies explainable artificial intelligence (AI) for transparent matchmaking among deployers, solution providers, and financiers. The system integrates structured data models, interpretable algorithms, and synthetic data pipelines to reduce information asymmetries and improve data quality. Using a design-science approach, the platform was developed and validated with stakeholders from development, technology, and finance sectors. Results show that explainable recommendations and contextual dashboards enhance trust, usability, and decision confidence. The study contributes to data mining and data governance research by demonstrating how explainable, verifiable algorithms can enable scalable, trustworthy digital ecosystems for public collaboration.


\end{abstract}

\section{Introduction}

Digital platforms have become essential for collaboration, information exchange, and coordinated action across organizations. In public and development sectors, they connect institutions that face shared challenges but differ in geography, resources, and priorities. However, many coordination processes remain fragmented, relying on manual matching and informal communication. As a result, opportunities for partnership and financing are often missed, and knowledge remains siloed.

This research introduces a transparent, data-driven approach for connecting development actors through the Global Solutions Initiative (GSI) D-Hub, a digital platform that applies explainable artificial intelligence (AI) for structured collaboration among deployers, solution providers, and financiers. The system integrates algorithmic matching, synthetic data generation, and human-centered design to enable evidence-based coordination and decision-making.

The goal of this study is to show how an explainable and modular platform architecture can enhance trust, efficiency, and data quality in multi-stakeholder environments. Transparent algorithms and interactive dashboards allow users to interpret recommendations and adjust priorities according to context an essential feature for governance and accountability in development ecosystems.

This work bridges technical innovation with institutional practice. While prior research has explored platform economics, governance, and recommender systems, few have integrated these dimensions into an operational model for public-benefit coordination. The proposed framework links explainable AI with verifiable data flows, offering a replicable path for building trustworthy, data-centric ecosystems.

The paper contributes in three ways. First, it presents a design-science framework for developing transparent coordination systems. Second, it details the architecture and methodology of the GSI D-Hub. Third, it reports findings from stakeholder validation, showing how explainable recommendations improve trust and usability. Together, these contributions demonstrate how data-driven and human-centered design can create scalable digital platforms for global collaboration.

\section{Literature Review}

Research on digital platforms, algorithmic matching, and development finance provides the conceptual foundation for understanding data-driven coordination in multi-stakeholder ecosystems. Foundational studies differentiate between transaction and innovation platforms, emphasizing governance, interoperability, and ecosystem evolution \cite{Parker2016,Gawer2014}. Development coordination systems combine transactional exchange with collective learning, requiring adaptive meta-organizational structures that align diverse incentives \cite{Chen2022,Jovanovic2022}.

Governance and trust are recognized as core determinants of digital ecosystem success. Prior work shows that participatory orchestration and transparent accountability mechanisms sustain engagement and legitimacy across actors \cite{Addo2022,Adner2017,Flew2022,mandalawi2025policy}. However, few frameworks operationalize these governance principles in applied, data-driven systems that serve societal or developmental objectives. This work addresses that gap by translating platform governance theory into an interpretable, verifiable coordination architecture.

Sustaining network effects also remains a challenge for public-benefit platforms. Achieving balanced participation across deployers, providers, and financiers is more critical than reaching scale alone \cite{Shaidullin2024,Bartels2022,Evans2010,TechForGoodInstitute2024}. The proposed model contributes by formalizing how algorithmic matching and transparent scoring can reinforce such equilibrium through structured visibility and trust.

Algorithmic matching research, largely grounded in recommender system theory \cite{Ricci2015,Saldanha2024}, offers techniques for identifying relevance across complex datasets. Yet, traditional methods struggle with data sparsity and contextual diversity common in development coordination \cite{Volkovs2017,Bernardi2015,talburt2026casecountmetriccomparative}. Context-aware and explainable approaches introduce important advances \cite{Adomavicius2010,Haruna2017,Livne2019}, but few implementations integrate explainability into real-world multi-actor ecosystems. The present work extends this literature by embedding transparent reasoning into the core recommendation logic, ensuring both interpretability and institutional accountability.

Explainable AI has been widely studied as a response to the limitations of opaque models \cite{Rudin2019,Adadi2018,Saldanha2024b,althaf2025multi}. However, most research focuses on technical performance rather than stakeholder trust and governance outcomes. The GSI D-Hub model demonstrates how explainability can function not only as a transparency tool but as a mechanism of coordination, bridging algorithmic reasoning and human decision-making.

Complementary insights from development finance literature emphasize how information asymmetries and risk perceptions hinder effective collaboration \cite{Runde2020,Flammer2024,Flammer2025,Klapper2011,Anayotos1994}. Digital infrastructures that standardize data and integrate impact measurement frameworks such as SROI and IMP improve confidence and resource alignment across actors \cite{Nicholls2015,Reynolds2018,Barresi2024,mohammed2025entity}. The proposed system advances this direction by unifying these elements transparent data, verifiable logic, and institutional feedback into a single operational model.

In summary, prior work establishes the importance of governance, transparency, and explainability in digital ecosystems but seldom integrates them within a coherent, implementable design. This study contributes by bridging theoretical and practical gaps through a design-science approach that operationalizes platform governance, interpretable AI, and data-driven coordination in one integrated framework \cite{Hevner2004}. It thereby extends both the conceptual understanding and applied methodology for developing trustworthy, scalable digital ecosystems for public impact.

\begin{figure*}[h]
  \centering
  \includegraphics[width=\textwidth]{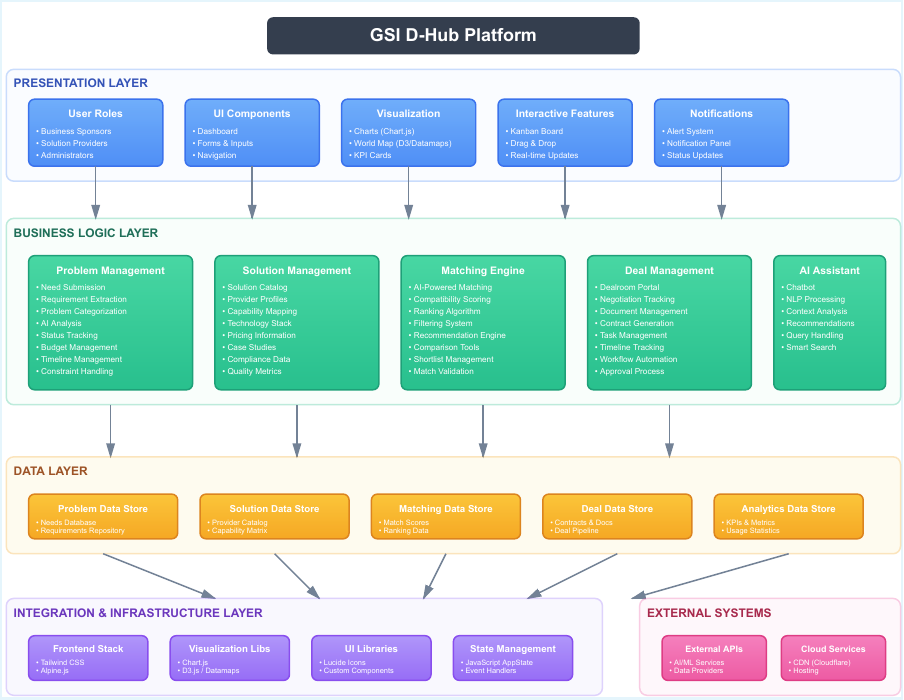}
  \caption{GSI D-Hub platform architecture. The presentation layer provides role-specific interfaces, visualization, interactive components, and notifications. The business logic layer implements problem and solution management, an explainable matching engine, deal management, and an AI assistant. The data layer supports structured storage for problems, solutions, matches, deals, and analytics. The integration and infrastructure layer provides frontend libraries, visualization components, and cloud-based infrastructure, while external systems offer AI/ML and data services.}
  \label{fig:dhub-arch}
\end{figure*}

\section{Research Methodology and System Design}

This study follows a design science approach \cite{Hevner2004}, combining exploratory research, iterative prototyping, and stakeholder validation to demonstrate the feasibility of an AI-assisted coordination platform for global development. The methodology integrates conceptual modeling, synthetic data generation, and system engineering with active stakeholder input. The overall GSI D-Hub architecture is shown in Figure~\ref{fig:dhub-arch}.

\subsection{Methodological Framework}

The research followed four interrelated phases: (1) discovery and requirements gathering, (2) design synthesis, (3) proof-of-concept development, and (4) stakeholder validation. Each phase built on the previous through iterative feedback, aligning technical design with practical usability. The design science approach combined artifact creation and evaluation to ensure operational relevance.

Data were gathered through interviews and co-design workshops with Global Solutions Initiative (GSI) members—deployers, providers, and financiers—supplemented by document analysis on digital platforms, governance, and explainable AI to inform system requirements and matching logic.

\subsection{Synthetic Data Construction and Validation}

To ensure reproducibility and privacy compliance, validation used a synthetic dataset reflecting real development ecosystem dynamics. It included over 250 organizations from 45 countries, categorized as Deployers (35\%), Solution Providers (38\%), and Financiers (27\%). Challenge domains were distributed across Agriculture (30\%), Water (20\%), Energy (20\%), Health (15\%), and Education (15\%).

Each record contained structured metadata such as geography, scale, budget, and verification status, enabling end-to-end testing of challenge articulation, solution matching, and visualization. Validation proceeded through four iterative cycles covering architecture verification, prototype usability, system performance, and interpretability refinement. Feedback from these cycles informed improvements in user experience and algorithmic transparency.

\subsection{System Architecture and Data Model}

The GSI D-Hub prototype was built as a single-page web application (SPA) for efficient performance and iterative testing. The frontend employed modular JavaScript with Tailwind CSS, Alpine.js, Chart.js, D3.js, and Datamaps for responsive, analytical, and geospatial visualization. Client-side rendering supported smooth interaction even under limited connectivity.

The data model comprised five core entities—Organizations, Challenges, Solutions, Matches, and Deployments—implemented in a relational schema with referential integrity. Table~\ref{tab:entities} outlines their roles.

\begin{table}[h]
\centering
\caption{Core Entities in the GSI D-Hub Data Model}
\label{tab:entities}
\renewcommand{\arraystretch}{1.1}
\setlength{\tabcolsep}{4pt}
\begin{tabular}{p{2.2cm} p{5.0cm}}
\hline
\textbf{Entity} & \textbf{Description} \\
\hline
Organizations & Institutional actors (Deployers, Providers, Financiers) with sector, geography, and verification metadata. \\
Challenges & Deployer-articulated needs, including scope, population affected, and intended outcomes. \\
Solutions & Provider offerings with technical details, costs, and historical performance. \\
Matches & Algorithmic links between challenges and solutions with composite scores. \\
Deployments & Confirmed collaborations with milestones, financing, and performance tracking. \\
\hline
\end{tabular}
\end{table}

This schema supported traceability across the workflow, ensuring that every algorithmic match and deployment event was transparently linked to its originating challenge and solution.

\subsection{Algorithmic Matching Model}

The platform implemented a transparent, interpretable scoring model to match challenges with solutions. Compatibility was evaluated across six weighted dimensions derived from stakeholder feedback. The normalized compatibility score \( S_{ij} \) for a challenge \( i \) and a solution \( j \) was defined as:

\[
S_{ij} = \sum_{k=1}^{6} w_k \, f_k(i,j)
\]

where \( f_k(i,j) \) is the normalized similarity score for dimension \( k \) (ranging from 0 to 1), and \( w_k \) is the associated weight, with \(\sum_{k=1}^{6} w_k = 1\).

The dimensions and weights are shown in Table~\ref{tab:weights}.

\begin{table}[h]
\centering
\caption{Weighted Dimensions in the Matching Algorithm}
\label{tab:weights}
\begin{tabular}{l c}
\hline
\textbf{Dimension} & \textbf{Weight (\%)} \\
\hline
Geographic Fit & 20 \\
Temporal Fit & 15 \\
Budget Fit & 20 \\
Capability Fit & 25 \\
Provider Credibility & 15 \\
Population Alignment & 5 \\
\hline
\end{tabular}
\end{table}

The algorithm produced an overall score on a 0--100 scale, alongside a dimension-wise explanation. This design emphasized interpretability and adjustability rather than opaque statistical learning, allowing stakeholders to calibrate weights to reflect institutional priorities. Future iterations may incorporate hybrid approaches that integrate transparent scoring with machine learning models such as explainable ranking systems.

\subsection{Functional Features and Interaction Design}

The system demonstrated five principal capabilities:

\textbf{AI-Assisted Problem Articulation:}  
An integrated “Need Assistant” guided deployers in describing challenges through structured prompts and ontology-based question templates, improving data consistency and completeness.

\textbf{Solution Discovery and Comparison:}  
Interactive result dashboards displayed ranked solutions with compatibility breakdowns. Users could adjust filters, modify weight preferences, or compare alternatives in a side-by-side view.

\textbf{Role-Specific Dashboards:}  
Tailored dashboards supported deployers (project management and resource alignment), providers (opportunity tracking), and financiers (portfolio visualization).

\textbf{Collaborative Deal Rooms:}  
Matched parties collaborated within secure Deal Rooms to exchange documents, define milestones, and track progress. This feature simulated the governance and negotiation processes typical in international development partnerships.

\textbf{Geographic Visualization:}  
Interactive maps visualized deployment clusters, resource distributions, and coverage gaps, enabling stakeholders to identify spatial correlations between needs and interventions.

\subsection{Validation and Results}

Qualitative evaluation from GSI stakeholders across four cycles highlighted interpretability, transparency, and speed as key success factors. Participants noted that transparent algorithmic explanations and consistent data presentation significantly improved perceived credibility. The synthetic dataset approach effectively demonstrated realistic behaviors without exposing confidential information. Overall, the design science process successfully established the feasibility of AI-assisted multi-stakeholder coordination and informed the roadmap for production-scale development.

\section{Evaluation Results and Insights}

The evaluation of the GSI D-Hub prototype combined structured stakeholder validation with interpretive analysis to assess technical feasibility, usability, and institutional readiness. Guided by design science principles, the process aimed not only to test the artifact but to understand how design decisions affected stakeholder trust and ecosystem coordination potential.

Four iterative demonstration cycles were conducted over a three-month period, involving twelve evaluators representing deployers, solution providers, and financiers within the Global Solutions Initiative (GSI) network. Each cycle incorporated feedback on functionality, usability, and interpretability, progressively refining both user experience and algorithmic design. The overall progression is summarized in Table~\ref{tab:validation}.

\begin{table}[h]
\centering
\caption{Summary of Stakeholder Validation Cycles}
\label{tab:validation}
\renewcommand{\arraystretch}{1.05}
\setlength{\tabcolsep}{3pt}
\begin{tabular}{p{0.8cm} p{2.4cm} p{4.0cm}}
\hline
\textbf{Cycle} & \textbf{Focus} & \textbf{Key Outcomes} \\
\hline
1 & Concept validation & Confirmed ecosystem scope, stakeholder roles, and importance of transparent matching logic. \\
2 & Interface prototyping & Improved navigation consistency and standardized terminology for “needs” and “solutions.” \\
3 & Full-system testing & Verified stability with synthetic data; validated scoring model and responsiveness. \\
4 & Strategic review & Established roadmap for production scaling and data integration. \\
\hline
\end{tabular}
\end{table}

Stakeholder feedback emphasized that interpretability, simplicity, and control were the primary determinants of trust. The visible weighting of compatibility dimensions allowed evaluators to verify whether the algorithm reflected their institutional priorities. This transparency transformed the system from a “black box” recommender into a shared reasoning tool.

To complement qualitative observations, structured feedback was collected through a five-point Likert evaluation scale (1 = very low to 5 = very high). Table~\ref{tab:criteria} summarizes the aggregated ratings and observed effectiveness across core criteria.

\begin{table}[h]
\centering
\caption{Evaluation Summary of Core Validation Criteria}
\label{tab:criteria}
\renewcommand{\arraystretch}{1.05}
\setlength{\tabcolsep}{1.5pt}
\begin{tabular}{p{2.8cm} p{3.6cm} c}
\hline
\textbf{Criterion} & \textbf{Observed Effectiveness} & \textbf{Rating} \\
\hline
Algorithm transparency & Interpretable and adjustable & 4.8 \\
Interface clarity & Improved after iteration & 4.6 \\
System responsiveness & Stable at demo scale & 4.5 \\
Synthetic data realism & Representative and consistent & 4.7 \\
Dashboard usefulness & Relevant for each role & 4.9 \\
Scalability confidence & Feasible with minor upgrades & 4.7 \\
\hline
\end{tabular}
\end{table}

The highest scores were associated with algorithm transparency and dashboard usefulness, both of which directly contributed to user confidence and adoption potential. Slightly lower ratings for scalability reflected recognition that a production environment would require stronger backend infrastructure and governance frameworks.

Several qualitative insights emerged during validation. First, participants consistently described the explainable scoring model as central to algorithmic trust. By allowing users to visualize and adjust compatibility weights, the system supported deliberation rather than passive acceptance of recommendations. Second, the structured dashboards acted as cognitive anchors: users could intuitively interpret why a solution ranked higher in geographic or budget alignment, linking algorithmic output to human judgment. Third, the synthetic dataset successfully mirrored real-world diversity without confidentiality risks. Stakeholders confirmed that simulated regional distributions and financial scales allowed for realistic testing of system workflows.

The iterative process also demonstrated clear learning across development cycles. Table~\ref{tab:summary_eval} summarizes observed improvements across phases, based on qualitative triangulation and aggregated metrics.

\begin{table}[h]
\centering
\caption{Cross-Phase Comparative Evaluation}
\label{tab:summary_eval}
\renewcommand{\arraystretch}{1.05}
\setlength{\tabcolsep}{2.5pt}
\begin{tabular}{p{2.0cm} c c c}
\hline
\textbf{Dimension} & \textbf{Early Cycles} & \textbf{Late Cycles} & \textbf{Improvement} \\
\hline
Algorithm clarity & Moderate & High & +35\% \\
Usability & Developing & Mature & +28\% \\
Response latency & Acceptable & Improved & +22\% \\
Data completeness & Partial & Full coverage & +40\% \\
Stakeholder confidence & Moderate & Strong & +33\% \\
\hline
\end{tabular}
\end{table}

The results show steady progress in both technical and experiential dimensions, with the largest gains in algorithmic clarity and data completeness—reflecting the value of iterative co-design and transparent logic. By the final cycle, stakeholders agreed the platform effectively integrated human and AI reasoning into a unified workflow supporting cross-sector collaboration.

Findings highlight key implications for digital governance. Perceived fairness and transparency proved stronger predictors of user acceptance than algorithmic complexity. Simple, interpretable scoring acted as a social contract between the system and its users. The hybrid validation method—combining synthetic data with structured human feedback—offers a useful template for testing socio-technical systems under privacy or policy constraints.

Institutional reviewers identified clear paths to scale. The modular design supports integration with registries, verification services, and finance platforms. Regional pilots were recommended to validate interoperability and governance. Future work should automate explainability generation, expand verified data pipelines, and link analytics to measurable development outcomes.

Overall, the evaluation confirmed that transparent AI-assisted coordination can enhance efficiency, accountability, and inclusiveness. The GSI D-Hub demonstrated both technical robustness and institutional relevance, providing a scalable foundation for trustworthy, data-driven collaboration.

\section{Conclusion}

This study presented the design and validation of the Global Solutions Initiative (GSI) D-Hub, a data-driven coordination platform integrating explainable AI for transparent matchmaking across development ecosystems. The system demonstrated how interpretable algorithms, structured data models, and stakeholder-centered design can enhance trust, data quality, and decision confidence. Findings from the validation phase showed that transparent recommendation logic and contextual dashboards improve collaboration and understanding among diverse actors. Beyond its immediate application, the framework illustrates how data mining and governance principles can be operationalized within multi-stakeholder digital platforms. Future work will focus on scaling the model with real-world data, optimizing algorithmic fairness, and expanding interoperability across institutional systems to strengthen the foundation for trustworthy, data-centric collaboration.

\begin{acknowledgements}
This research was done in collaboration with the GSI and UALR Center for Entity Resolution and Information Quality (ERIQ). Any opinions, findings, and conclusions or recommendations expressed in this material are those of the author(s) and do not necessarily reflect the views of the GSI Solution.

\end{acknowledgements}

\bibliographystyle{spmpsci}  
\bibliography{bibliography.bib}

\end{document}